\documentclass{article}[12pts]
\usepackage{setspace}
\usepackage[utf8]{inputenc}
\usepackage[margin=1in]{geometry}
\usepackage{amsmath}
\usepackage{amssymb}
\usepackage{amsfonts} 
\usepackage{bm}
\usepackage{listings}
\usepackage{graphicx}
\usepackage[english]{babel}
\usepackage{inputenc}
\usepackage{caption}
\usepackage{array,multirow}
\usepackage{subcaption}
\usepackage{xcolor}
\usepackage{hyperref}
\usepackage{float}
\usepackage{tabularx}
\usepackage{adjustbox}
\usepackage[version=3]{mhchem}
\usepackage{siunitx}

\usepackage[capitalize]{cleveref}
\usepackage{authblk}  
\title{A Bayesian approach for unadjudicated events in cardiovascular disease cohort studies}
\author[1] {Mirajul Islam}
\author[2]{Michael J. Daniels}
\author[3]{Donald Lloyd-Jones}
\author[4]{Juned Siddique}
\affil[1]{PhD Candidate, Department of Statistics, University of Florida}
\affil[2]{Professor, Department of Statistics, University of Florida}
\affil[3] {Department of Preventive Medicine, Northwestern University, Chicago, IL (D.M.L.-J.)}
\affil[4]{Professor, Department of Preventive Medicine, Northwestern University Feinberg School of Medicine, Chicago}
\date{\today}
\begin{document}
\maketitle
\begin{abstract}
An important issue in joint modelling for outcomes and longitudinal risk factors in cohort studies is to have an accurate assessment of events. Events determined based on ICD-9 codes can be very inaccurate, in particular for cardiovascular disease (CVD) where ICD-9 codes may overestimate the frequency of CVD. Motivated by the lack of adjudicated events in the Established Populations for Epidemiologic Studies of the Elderly (EPESE) cohort, we develop methods that use a related cohort Atherosclerosis Risk in Communities (ARIC), with both ICD-9 code events and adjudicated events, to create a posterior predictive distribution of adjudicated events. The methods are based on the construction of flexible Bayesian joint models combined with a Bayesian additive regression trees to directly address the ICD-9 misclassification. We assessed the performance of our approach by simulation study and applied to ARIC data.
\end{abstract}

\section{Introduction}
The International Classification of Diseases (ICD) is a diagnostic tool used to classify causes of death from death certificates. These codes are widely used in epidemiological research to assess public health and inform policymakers about health care and research. Although the use of ICD code is convenient and inexpensive compared to performing an adjudication process, ICD coding has a limited ability to accurately identify specific diagnoses \cite{benesch1997inaccuracy,goldstein1998accuracy,rinaldi2003accuracy,golomb2006accuracy}. Khan et al. \cite{khan2018limited} found that a large academic health care system had less than 50\% accuracy at classifying a true congenital heart disease using ICD‐9, Clinical Modification (CM) administrative codes among patients with various types of disease lesions. Rodriguez et al. \cite{rodriguez2022well} assessed the performance of ICD-9-CM codes to predict congenital heart defects and proposed the need to develop algorithms to improve the identification. Analyses, performed using unadjudicated events as confirmed events, are subject to inherent bias, and overestimating rates, resulting in invalid inference \cite{cook2004analysis}.\\
 In a cohort study, when the primary interest is the time to event such as  CVD death, or myocardial infarction and only unadjudicated events are recorded, it is necessary to review by a blinded event classification committee (ECC) \cite{seltzer2015centralized}. ECC determines whether a reported event using ICD is a true event or not; the resulting event is an adjudicated event. This process removes misclassification, but is expensive and time consuming. Another way of avoiding misclassification is to estimate the unadjudicated events by building a model from available adjudicated events given unadjudicated events \cite{cook2004analysis}. Studies have been conducted to obtain more accurate estimates of the parameters of interest using statistical tools to estimate missing failure indicators when a subset of them is not available \cite{magaret2008incorporating,dodd2011audit}.\\
 With a goal of estimating the effects of covariates on the hazard of events, several papers proposed approaches for inference with survival data with missing censoring indicators using parametric, non-parametric and semiparametric hazard models \cite{subramanian2000efficient,gijbels2007non,hyun2012proportional,brownstein2015parameter,goetghebeur1995analysis,zhou2003additive}.  Cook \cite{cook2000adjusting} proposed methods for analyzing survival data with incompletely adjudicated events, which were further extended by Cook and Kosorok \cite{cook2004analysis} to develop inference procedures for time-to-event data that yield consistent and asymptotically normal estimators by weighting observations based on their probability of being true cases, with standard errors obtained via bootstrapping. Qiu et al \cite{qiu2015kernel,qiu2021quantile} introduced a kernel-assisted imputation estimating method. A reweighting method based on augmented inverse probability of failure indicators is developed and asymptotic properties of the estimators are discussed in \cite{chen2018reweighted}. A recent study proposed a shared parameter joint model using cumulative incidence function with an extension to account for failure cause misclassification using double sampling \cite{10.1093/biostatistics/kxac043}. We propose a new flexible Bayesian machine learning approach for inference to determine whether an unadjudicated event is a true event from the cohort that has both adjudicated and unadjudicated events. This method uses i) features of longitudinal risk factors with baseline covariates to determine true event taking into account uncertainty, and ii)  a joint modeling framework that accommodates incomplete adjudication of time-to-event outcomes by integrating longitudinal biomarkers and latent event status under a probabilistic adjudication mechanism (unlike previous work).\\
Our motivating study is a large cohort study, the Established Populations for Epidemiologic Studies of the Elderly (EPESE). This study only contains CVD death status based on the ICD-9 CM code. Adjudicated CVD events are not available in this study. Our objective is to: i) build a model for whether an unadjudicated event is a true event using data from Atherosclerosis Risk in Communities (ARIC) study that has both adjudicated and unadjudicated CVD death events that uses longitudinal risk factor features, and ii) propose an approach to get more precise estimates of the hazard parameters in joint model. \\
This paper is organized as follows. Section 2 describes the cohort data and variables. Section 3 introduces the specification of the  novel adjudication approach with descriptions of the models and details on inference. Section 4 and Section 5 present the simulation results and application on ARIC data including the performance of our approach. Finally, in Section 6, we end with a discussion.
\section{Motivating study}
EPESE study collected information on chronic conditions, death, disabilities, and institutionalization of elderly people living in four communities: New Haven, Connecticut; East Boston,  Massachusetts; five counties in north central North Carolina; and two counties in Iowa during 1981-1993 \cite{cornoni1986established,huntley1993established}. EPESE includes more than thirteen thousand participants, aged 65 years and older. The study initially collected data during a (baseline) household interview. Then, it continued to collect information using surveillance of morbidity and mortality. Participants were followed every year to collect data on cause of death and factors related to nursing home admissions and hospitalization. Information from state/hospital records and death certificates are also collected. We remove the participants with self-reported CVD history at the baseline interview from the EPESE data.\\
The ARIC study, a population-based, prospective cohort study consists of over 15,000 individuals' aged 45-64 years at the beginning of the study \cite{atherosclerosis1989atherosclerosis}. Baseline information was collected including medical history, physical activity, medication use, and diet during 1987-89. Data was collected from participants at 6 visits during 1990-2019. In ARIC, total CVD death based on ICD-9 code is 2748 (Table (\ref{Tab:1})). Among these only 1597 are truly CVD death according to adjudication i.e., the missclassification rate is 42\%. The sensitivity and specificity of ICD-9 are 58\% and 85\%, respectively.
\begin{table}[H]
   \centering
		\caption{Misclassification in ARIC data}
    \label{Tab:1}
		\begin{tabular}{c|ccc}
			\hline
&&\multicolumn{2}{c}{ICD-9}\\
\hline
   &&Death from CVD&Death from NOT CVD\\
     \multirow{2}{*}{Adjudication}&Death from CVD&1597 (58\%)&841 (15\%)\\
&Death from NOT CVD&1151 (42\%)&4852 (85\%)\\
   			\hline
		\end{tabular}
	\end{table}
We only consider ARIC individuals' information from age 55 years and older given the age range of EPESE. The variables, used in this study are  baseline body mass index (BMI), race (black vs white), sex  and education level ($<$ High School (LH), High School, $>$ High School (AH)). The used longitudinal risk factors are systolic blood pressure (SBP) and diastolic blood pressure (DBP), glucose level (GLUCOSE), and total cholesterol (TOTCHL). The non-fatal event variables include the number of heart failure incidents, myocardial infarctions, and strokes up to the last visit (all based on ICD-9 codes). 
Given the lack of ICD-9 event predictors and the few longitudinal risk factors in EPESE data we were unable to apply our approach to EPESE. As such, we divide the ARIC data into two parts: training dataset (A) (75\% of ARIC) and test dataset (B) (25\% of ARIC).
\section{Approach to adjust for unadjudicated events}
Since misclassification only occurs among dead individuals, we consider only dead individuals from both datasets to adjust for the unadjudicated events. First, we use risk factor models to compute the posterior distribution of the longitudinal features. Then we fit Bayesian Additive Regression Tree (BART) models to dataset A to estimate posterior probabilities of CVD death for the dataset B including the features from the risk factor model as covariates. Finally, we estimate survival parameters fitting a joint model on the full (both dead and alive) B dataset.
\subsection{Notation}
Let $D_i$ and $D_i^*$ be the ages at unadjudicated death from CVD (based on ICD) and adjudicated death from CVD, respectively. Also, let $\Delta_i = I\{ICD(9)=CVD\}$ and $C_i=I\{\mbox{Death due to CVD}\}$ be CVD death indicator based on ICD-9 code and death status from CVD after adjudication, respectively.
	
\subsection{Risk factor model for A and B}\label{RF}
Let $y_{gi}(t)$ be the $g$th $(g=1,2,\ldots,G)$ longitudinal outcome for $i$th individual at age $t$. Two separate mixed-effect models are fitted one for $\Delta_i=1$ and one for $\Delta_i=0$ i.e., for ICD9$=$CVD and ICD9$\neq$CVD, respectively. Given  non-linear longitudinal trajectories (see Figure \ref{Fig_traj}), we use a second order shifted Legendre orthogonal polynomial for the effect of age in both fixed and random effects,
$$Y_{gi}(t) |\Delta_i=\mu_{gi}(\Delta_i,t)+\epsilon_{gi}^{\Delta_i}(t)$$
where $\mu_{gi}( \Delta_i,t)=\beta_{i0}+\beta_{i1}P_1(t)+\beta_{i2} P_2(t)+ \beta_3{SEX}_i+\beta_4 {RACE}_i+\beta_5 {LH}_i+\beta_6 {AH}_i; P_1(t)=\frac{2t}{T}-1,P_2(t)=\frac{1}{2}(3{(\frac{2t}{T}-1)}^2-1)$ and $ \epsilon_{gi}^{\Delta_i}(t)\sim N(0,\sigma^2_{g}).$\\
Instead of restricting the distribution of the random effects to be normal, we assume a more flexible, nonparametric structure of the random effects using a Dirichlet process mixture (DPM) which we formulate in the following manner:
$$\bm{b}_i=(\beta_{i0},\beta_{i1},\beta_{i2})^T\sim N(\mu_i,\Sigma_i),~~(\mu_i,\Sigma_i)\sim H,~~ H\sim DP(\alpha, H_0),~~H_0\sim Normal\times IW.$$
where DP($\cdot$) denotes a Dirichlet process prior, $\alpha$ is a scalar precision parameter. DP can also be written using the stick-breaking construction \cite{sethuraman1994constructive}. The truncated version of DPM, proposed by Ishwaran and James \cite{ishwaran2001gibbs}, uses the stick-breaking construction and can be expressed as
$$\bm{b}_i|z=k\sim MVN(\bm{\mu_{k}},\Sigma_{k}), k=1,2,...,K$$ $$z\sim Mult(K,\bm{\pi})$$
$$\pi_1=V_1, \pi_K=1-\sum_{k=1}^{K-1}\pi_k $$
				$$\pi_k=V_k\prod_{l=2}^{k-1}(1-V_l), k=2,...,K-1$$
				$$\alpha\sim Gamma(a,b); V_k|\alpha\sim Beta(1,\alpha), k=1,...,K-1$$
				$$\bm{\mu_k}\sim MVN(\bm{\mu_0},D); \Sigma_k\sim invWishart(R_1,c).$$
The truncation number K in the DPM was chosen following \cite{ohlssen2007flexible}.We choose hyper prior parameters in the DPM following Taddy's method \cite{taddy2010bayesian}. \\
From the risk factor model, we compute the following features
	\begin{equation}
  \label{feature}
  \begin{gathered}[b]
		f_{g1}\{\mu_i(\Delta_i),t\}=\mu_{ig}(\Delta_i,t)~~~~~~~~[\text{value}]\\
		f_{g2}\{\mu_i(\Delta_i),t\}=\frac{d\mu_{ig}(\Delta_i,t)}{dt}~~~~~~[\text{slope}]\\
		f_{g3}\{\mu_i(\Delta_i),t\}=\int_{t_0}^t\mu_{ig}(\Delta_i,s)ds~[\text{area}]
 \end{gathered}
\end{equation}
where $f_{g1}(.)$ is the current value of the time-varying risk factor, $f_{g2}(.)$ is the slope of the true trajectory of that risk factor, and $f_{g3}(.)$ denotes the area feature, i.e., cumulative effects of the time-varying risk factor, respectively. 
\subsection{BART probit model for dataset A}
We fit the BART probit model \cite{chipman2010bart} using SoftBart package in R \cite{linero2022softbart}. The BART probit model is fitted separately for $\Delta_i=1$ and $\Delta_i=0$. Let $X_i$ be a covariate vector that includes baseline covariates, indicators of unadjudicated non-fatal events, and age at unadjudicated death for the $i$th individual. The BART probit model is,
\begin{eqnarray}\label{BART}
P(C_i=1 | \Delta_i,X_i, f(\mu_i(\Delta_i)))&=&\Phi\Big(\sum_{j=1}^m g(\Delta_i,X_i, f(\mu_i(\Delta_i)); T_j,M_j)\Big), \label{eq:bart1}
\end {eqnarray}
where  $T_j$ is a binary regression tree with associated terminal node parameters $M_j=(\mu_{j1},\ldots,\mu_{jb_j})$, $g(.)$ is a function that assigns $\mu_{ji}$ in $M_j$ to $(X_i,f(\mu_i(\Delta_i)))$ and $\Phi(.)$ is the cdf of standard normal distribution. The model (\ref{BART}) uses data augmentation \cite {albert1993bayesian} and
implicitly assumes a standard deviation 1. A regularization prior is specified on $(T_1,M_1),\ldots ,(T_m,M_m)$. This prior can be factored and specified as in \cite{chipman2010bart},
$$p((T_1,M_1),\ldots(T_m,M_m))=\prod_j\Big[p(T_j)\prod_ip(\mu_{ij}|T_j)\Big].$$

\subsection{Joint model for dataset B}
Let $x_{ig}$ be the design vector  of the $g$th risk factor for the fixed-effect regression coefficients ($\bm{\beta}_g$) and  $z_{ig}$ be corresponding design vector for the random effects ($\bm{b}_{g}$). Then, for the $i$th individual, the $g$th longitudinal sub-model is
\begin{eqnarray*}
    y_{gi}(t)&=&\mu_{gi}(t)+\epsilon_{gi}(t); g=1,2,...,G;~i=1,2,\ldots,n,
\end{eqnarray*}
 where $\mu_{gi}(t)$ has same structure as mean and random effects in \ref{RF}, $\bm{\beta_g}=(\beta_{g0},\beta_{g1},\ldots,\beta_{gp})^T,\epsilon_{ig}\sim N(0,\sigma_g^2).$
To account for the correlation among the $G$ longitudinal
outcomes and also correlation within each longitudinal outcome, we assume a multivariate normal distribution for the corresponding random effects as follows
\begin{equation}\label{eq:2}
	\bm{b}_i = (b_{1i}^T,b_{2i}^T,\ldots,b_{Gi}^T)\sim N(\mathbf{0}, D).
\end{equation}
 For the survival outcome, let the observed time $T_i=\text{min}~(T_i^*,C_i^*)$ where $T_i^*$ and $C_i^*$ be the event time and censoring time, respectively. Assuming the risk of the event depends on the features of the risk factors and baseline covariates ($w_i$), the survival sub-model is written as
\begin{equation}\label{eq:1}
\lambda_i(t,\bm{\theta_s}) = \lambda_0(t) \exp[w_i^T\bm{\gamma}+\sum_{g=1}^{G}\sum_{j=1}^{J}\alpha_{gj}f_{gj}\{\Psi_{gi}(t\}]
\end{equation}
	where $\Psi_{gi}(t)=\{\mu_{gi}(s),0\leq s\leq t\}$ is the history of the $g$th true unobserved longitudinal process
	up to time point t, $\bm{\theta_s}$ is the parameter vector for the survival outcomes, $\bm{\gamma}$ is the regression coefficients of the baseline covariates, and $\alpha_{gj}$  is a set of parameters that link the longitudinal features with survival outcome. We specify the baseline hazard function using B-splines \cite{andrinopoulou2016bayesian}.\\
To link the survival process with longitudinal outcomes, we used the three features given in (\ref{feature}).
\subsubsection{Prior specification}
It is important to note that the features within a risk factor are often correlated.  For example, if there is little change over time, the current value and cumulative features will be strongly correlated. For this reason, there are typically not more than one or two important features for a risk factor. To account for the correlation among features within a risk factor, we use BSGS-D priors  \cite{islam2024bayesian} on $\alpha_g$ except replacing the spike with a narrow population model that concentrates around zero and a slab with wider normal population model \cite{george1993variable}. We do this as Hamiltonian Monte Carlo (HMC) in Stan does not allow updates of discrete parameters of the log posterior density. We reparametrize the coefficient vectors that link the longitudinal and survival process in (\ref{eq:1}) as follows: 
	$$\bm{\alpha}_g = \bm{V}_g^{\frac{1}{2}}\bm{d_g}, \text{where}~ \bm{V}_g^{\frac{1}{2}}=diag\{\tau_{g1},\tau_{g2},\ldots,\tau_{gJ}\}, \tau_{gj}\geq 0.$$
 Following \cite{george1993variable}, the spike and slab prior is
 	$$\bm{d}_g\overset{ind}{\sim}(1-\pi_{g}) N_{J}(\bm{0}, \tau_g^2\bm{I}_{J})+\pi_g N(0,c_g^2\tau_g^2\bm{I}_{J}),~~~ g = 1,2,\ldots,G.$$
  To choose $\tau_g$ and $c_g$, George and McCulloch \cite{george1993variable} suggested a semiautomatic approach by considering the intersection point and relative heights at 0 of the marginal densities $$d_{gj}|\sigma_{d_{gj}}\pi_g=0\sim N(0,\sigma_{d_{gj}}^2+\tau_g^2)$$
$$d_{gj}|\sigma_{d_{gj}}\pi_g=1\sim N(0,\sigma_{d_{gj}}^2+c_g^2\tau_g^2).$$
For this form, their choices were $(\sigma_{d_{gj}}/\tau_g,c_g)=(1,5),(1,10),(10,100)$ etc. Here, we assume $\sigma_{d_{gj}}^2+c_g^2\tau_g^2=1,\sigma_{d_{gj}}/\tau_g=1~\text{and}~c_g=10$, to obtain $\tau_0=\sqrt{\sigma_{d_{gj}}^2+\tau_g^2}=0.1407.$ Finally, the spike and slab prior becomes
 	$$d_{gj}|\pi_g\sim(1-\pi_g)N(0,\tau_0^2)+\pi_g N(0,1)$$
  $$\tau_{gj}\overset{ind}{\sim}(1-\pi_{gj})N^+(0,\tau_0^2s^2)+\pi_{gj}N^+(0,s^2),$$
	where $s^2\sim \text{Inv-Gamma}(a,b)$  and $\pi_g\sim \text{Beta}(c,d)$.
Probabilities $\pi_{gj}$ are calculated exactly as in the BSGS-D prior to taking into account the correlation among features within a risk factor (details in the supplement). We adopt a non-centered parameterization for the random effects covariance matrix using a Cholesky decomposition,
$$\bm{b}_i=L\bm{z}_i,~~\bm{z}_i\sim N_q(\bm{0},I),~~ L=\Sigma_0L_{c};~~\Sigma_0=\text{Diag}(\sigma_1,\ldots,\sigma_{q})$$
$$\sigma_j\sim N^+(0,1); j=1,2,\ldots,q,~~L_{c}\sim \text{LKJ}(1),$$
where LKJ (named after Lewandowski, Kurowicka, and Joe) is a prior distribution that with parameter equal to one defines uniform distribution over all valid correlation matrices.
This parameterization ensures that $D=\Sigma_0 R\Sigma_0$ in \ref{eq:2} where $R=L_{c}L_{c}^T$.We assume a diffuse normal priors for the longitudinal regression parameters$(\bm{\beta_g})$ and the baseline survival parameters $\bm{\gamma}$.	
\subsection{Transportability assumption}
To use our approach for a different cohort like EPESE, we also need a transportability assumption. Let $\pi^{E}=P(C_i^E=1 | \Delta_i,X_i, f^E(\mu_i(\Delta_i)))$ and $\pi^{A}=P(C_i^A=1 | \Delta_i,X_i, f^A(\mu_i(\Delta_i)))$. The transportability assumption is
$$\pi^{E}= \pi^{A}.$$
That is, the regressions would be the same even if the distributions of the key characteristics [the features, covariates and CVD death indicator by ICD-9] were different. It assumes that the relationships between variables and the CVD death indicators, $C_i$ are sufficiently similar between the two cohorts, allowing one to ``transport'' the results from one cohort to another.
\subsection{Algorithm for inference}
Let $n_B$ and $N_B$ be the number of only dead and total (dead and alive) individuals in dataset B, respectively. 
\subsubsection{Adjustment for unadjudicated events}
Conduct the following steps among only dead individuals separately for ICD9 CVD event ($\Delta_i=1$) and ICD9 NOT CVD event ($\Delta_i=0$).
\begin{description}
    \item  Step 1: Fit risk factor models and obtain posterior distribution of the features both for A ($F^A$) and B datasets ($F^B$). 
    \item  Step 2: Using $m=1,2,\ldots,M$ posterior samples from posterior distribution of the features for both datasets, do the followings:
    \begin{itemize}
        \item fit a BART model ($C_i^A|F^A_{im},X_i,\Delta_i;\theta_c$) using the posterior features and baseline covariates from A 
        \item from the fitted model, compute  $k=1,2,\ldots,K$ posterior samples of probabilities for individuals from B using posterior features and baseline covariates from B i.e.,
         \begin{eqnarray}
     \hat{w}_{imk}=\hat{P}_{imk}(C_{i}^B=1|F^B_{im},X_i,\Delta_i)&=&\Phi\Big(\hat{G}(F^B_{im},X_i,\Delta_i)\Big). \label{eq:predbart1}
\end {eqnarray}
        \item  for $k=1,2,\ldots,K$, generate $$U_{imkl}\sim Uniform(0,1); l=1,2,\ldots,L.$$ and create adjudicated CVD death status $(A_{imkl})$ as\\       
        \begin{equation*}
  A_{imkl}=
    \begin{cases}
      1 & \text{if}~U_{imkl}<\hat{w}_{imk}\\
      0 & \text{otherwise}.
    \end{cases}       
\end{equation*}
        \end{itemize}
 \end{description}
  From these steps, we have the $M*K*L$ sets of adjudicated CVD death indicators for each dead individual in dataset B.
 \subsubsection{Fitting joint model}
  \begin{itemize}
   \renewcommand\labelitemi{}
  \item Step 3: First, create an CVD death status indicator for all (both dead and alive) individuals in dataset B as, 
        \begin{equation*}
  A_{imkl}^* =
    \begin{cases}
      A_{imkl} & \text{if $i$th individual dead}\\
      0 & \text{if $i$th individual alive}.
    \end{cases}       
\end{equation*}
        \item  Step 4: Now, fit a joint model on each of the $M*K*L$ sets of the full B dataset using $A_{imkl}^*$. This results in $N=M*K*L$ posterior estimates for $p$th parameter,$\theta_p$. \\
       
    \item  Step 5: Compute the Monte Carlo standard error (MCSE) of the estimated survival parameters (${\theta}_{p}$) from the posterior estimates as, $$\hat{\hat{\theta}}_p=\frac{1}{N}\sum_{i=1}^{N}\hat{\theta}_{pi};~\text{and~MCSE}(\hat{\theta_p})=\sqrt{\frac{1}{N}\frac{\sum_{i=1}^{N}(\hat{\theta}_{pi}-\hat{\hat{\theta}}_p)^2}{N}};~ p=1,2,\ldots,P.$$

     \item Step 6: Compute $95\%$ credible intervals of the survival parameters (${\theta}_{p}$) as, 
     $$\hat{\hat{\theta}}_p\pm 1.96\times\text{MCSE}(\hat{\theta_p})$$
\end{itemize}
We choose an $N$ so that MCSE is practically negligible. Details are provided in the next subsection. 
\subsubsection{Choice of $M, K \& L$}
 To choose $M,K,L$ for a given $\epsilon_{\hat{\theta}_p}$, we compute the total variability for each $\hat{\theta}_p$ (say $SST_{\hat{\theta}_p}$) for an initial choose of M,K,L; we recommend $M=K=L=3$. We then compute $N_{\hat{\theta}_p}=M_{\hat{\theta}_p}\times K_{\hat{\theta}_p}\times L_{\hat{\theta}_p}$ using $SST_{\hat{\theta}_p}$ and $\epsilon_{\hat{\theta}_p}$ by solving the following inequality
	$$\sqrt{\frac{SST_{\hat{\theta}_p}/N_{\hat{\theta}_p}}{N_{\hat{\theta}_p}}}\leq \epsilon_{\hat{\theta}_p}.$$
We choose the value of $\epsilon_{\hat{\theta}_p}$ based on the magnitude of the parameter estimate $\hat{\theta}_p$. For example, if the estimates of the parameters 1 and 2 are 0.25 and 0.025, respectively, we suggest choosing $\epsilon_{\hat{\theta}_1}=0.01$ and $\epsilon_{\hat{\theta}_2}=0.001$ corresponding to values rounded to the second significant digit of the parameter estimates.  Finally, $N=\max\{N_{\hat{\theta}_p}\}$ and we choose the values of $M, K$, and $L$ accordingly. This procedure can be very efficient given high performance computing,
\begin{itemize}
\item Computing $M$ posterior features for both ARIC and EPESE requires time of fitting one mixed model, with computational time denoted by $T^F$.
\item Fitting the BART model on ARIC using the $M$ posterior features requires time $T^{\text{fit}}$.
\item Computing posterior weights for EPESE involves predicting from the fitted BART model, requiring time $T^{\text{pred}}$.
\item Running $M \times K \times L$ joint models requires the time to fit one joint model, denoted by $T^{\text{JM}}$.
\end{itemize}
Therefore, given values of $M$, $K$, and $L$, and assuming unlimited computational cores, the total computation time is:~ $T^F+T^{fit}+T^{pred}+T^{JM}$. 
\section{Simulation}
We conducted a simulation study to examine the performance of our approach. We simulated data similar to ARIC, with 1200 subjects and a maximum of 5 follow-up visits by fitting the below mixed model to ARIC to get true parameter estimates. The steps are: \\
 We sample 1200 (800 events and 400 censored by adjudication) dead individuals info from ARIC and fit the following multivariate mixed model on average blood pressure, glucose and TOTCHL.
    \begin{equation}\label{sim_feat}
    y_{ig}(a_{i\ell})=\beta_{g0}+\sum_{k=1}^2\beta_{k,g}P_k(a_{i\ell})+b_{i0,g}+\sum_{k=1}^2b_{ik,g}P_k(a_{i\ell}) +\epsilon_{ig}(a_{i\ell}),
    \end{equation}
	where $P_1(a_{i\ell})=(\frac{2a_{i\ell}}{a_{max}}-1); P_2(a_{i\ell})=\frac{1}{2}( 3(\frac{2a_{i\ell}}{a_{max}}-1)^2-1); g=1,2,3;~i=1,2,\ldots,1200,\epsilon_{ig}\sim N(0,\sigma_g^2)~\text{and}~\bm{b_i}\sim N(\bm{0},\bm{D})$ and find the `true' parameters $\beta,\sigma, D$ using the posterior means.
Then, using the true values of $\beta,\sigma, D$ we calculate current value features ($V_g$). Finally, we fit a Cox model using CVD death by adjudication as the event as,
\begin{eqnarray}\label{eq:surv}
				\lambda_i(t)& =& \lambda_0(t) \exp\Big[\gamma_0{Base age}+\gamma_1{BMI}_i+\gamma_2{LH}_i+\gamma_3{AH}_i+\gamma_4{Sex}_i+\gamma_5{Race}_i+\sum_{g=1}^3\alpha_gV_g\Big]
		\end{eqnarray}
   and compute the ``true" parameters $\gamma$ and $\alpha$ from the fitted model above based on posterior means.\\
We simulate the baseline (centered) age uniformly from 0 to 19 and use it as a covariate in the hazard model. Following ARIC, the centered ages at the visits were created by adding 3-year intervals after setting the baseline age 0. Using the values of the true longitudinal parameters obtained previously, we calculate longitudinal risk factors. The baseline risk was simulated from a Weibull distribution $\lambda_0(t) =at^{a-1}$ with $a = 1.75$. An exponential censoring distribution with mean 8 was applied. To estimate the parameter coefficients underlying the simulation models, we fitted joint models.\\
Next we fit the following Bayesian logistic regression on CVD death by ICD9 from ARIC ($\Delta_i$) including baseline variables, features and non-terminal events as covariates ($X$) separately for CVD death ($C_i=1$) and NOT CVD death by adjudication ($C_i=0$) (recall our approach predicts $C_i|\Delta_i$) on each simulated dataset as, 
    \begin{eqnarray*}
     \text{logit}(P(\Delta_i=1 | C_i,X_i, f(\mu_i(C_i))))&=&\delta^TX_i,
     \label{eq:logit}
 \end{eqnarray*}\label{Bayes-logit}
   and generate true unadjudicated CVD death from above fitted Bayesian logistic regression as follows:
    $$\text{CVD-UNADJ}=ifelse(\hat{P}>U,1,0); U\sim Unif(0,1),$$
    where $\hat{P}$ is the posterior probabilities obtained from Bayesian logistic regression.\\
We partitioned the simulated data set into a training and test dataset. For simplicity, we fit our proposed approach for $M=K=L=3$. The posterior distribution was sampled for 500 replicated datasets and the results were compared in terms of bias, root mean squared error (RMSE) and coverage probabilities of the hazard parameters ($\gamma,\alpha$). For each simulated dataset, we fit joint model on simulated adjudicated CVD deaths and another joint model on simulated unadjudicated CVD deaths. We also implement our approach as described in Section 3. The longitudinal part and survival part of joint model were as given in \ref{sim_feat} and \ref{eq:surv}. 
\subsection{Results}
The performance of the three approaches—using adjudicated CVD, unadjudicated CVD, and our proposed method is evaluated in terms of bias, and RMSE for the survival parameters. These metrics are presented in Tables \ref{tab:bias}, and \ref{tab:MSE}, respectively.\\\\
The bias comparison reveals that for most parameters, all three approaches, adjudicated, unadjudicated, and our proposed method, exhibit bias in the same direction, whether under- or overestimated. An exception is observed for the coefficient corresponding to individuals with lower than higher secondary education, where the unadjudicated approach yields a positive bias, in contrast to the negative bias observed under both the adjudicated analysis and our method.\\
Overall, the adjudicated data consistently result in the smallest bias and MSE. Importantly, our approach substantially reduces both bias and MSE relative to the unadjudicated analysis across nearly all parameters, indicating its effectiveness in correcting for misclassification due to incomplete adjudication.
\begin{table}[H]
	\centering
	\caption{Bias comparison for the survival parameters}
	\begin{tabular}{c|c|c|c|c|c|c}
    \hline
    Variable&\multicolumn{3}{c|}{Bias}&\multicolumn{3}{c}{Absolute bias}\\
    \hline
   & Adjudication& Unadjudication& Our approach&  Adjudication& Unadjudication& Our approach\\
    \hline
    Value-BP&0.020&0.028&0.027&0.020&0.028&0.027\\
    Value-GLUC&-0.267&-0.332&-0.132&0.267&0.332&0.132\\
     Value-CHL&-0.014&-0.018&-0.018&0.014&0.018&0.018\\
      Base-age&-0.015&-0.025&-0.021&0.015&0.025&0.021\\
       BMI&-0.006&-0.006&-0.010&0.006&0.006&0.010\\
        LH&-0.030&0.102&-0.019&0.030&0.102&0.019\\
         AH&0.127&0.261&0.177&0.127&0.261&0.177\\
          Sex&-0.136&-0.274&-0.190&0.136&0.274&0.190\\
          Race&0.133&0.225&0.187&0.133&0.225&0.187\\
          \hline
	\end{tabular}
	\label{tab:bias}
\end{table}
\begin{table}[H]
	\centering
	\caption{MSE comparison for the survival parameters}
	\begin{tabular}{c|c|c|c|c|c|c|c}
    \hline
    &&\multicolumn{2}{c|}{Adjudication}&\multicolumn{2}{c|}{Unadjudication}&\multicolumn{2}{c}{Our approach}\\
    \hline
    Variable&Truevalues&Mean& RMSE&Mean& RMSE& Mean& RMSE\\
    \hline
    Value-BP&-0.050&-0.030&0.021&-0.022&0.028&-0.023&0.028\\
    Value-GLUC&0.000&-0.267&3.179&-0.332&6.084&-0.132&1.307\\
     Value-CHL&0.032&0.018&0.014&0.014&0.019&0.014&0.019\\
      Base-age&0.050&0.035&0.022&0.025&0.029&0.029&0.025\\
       BMI&0.020&0.014&0.014&0.014&0.017&0.010&0.015\\
        LH&0.035&0.005&0.217&0.137&0.273&0.016&0.191\\
         AH&-0.489&-0.362&0.259&-0.228&0.353&-0.312&0.267\\
          Sex&0.474&0.338&0.223&0.200&0.325&0.284&0.244\\
          Race&-0.470&-0.337&0.284&-0.245&0.353&-0.283&0.250\\
          \hline
	\end{tabular}
	\label{tab:MSE}
\end{table}
Table \ref{tab:prob} presents the coverage probabilities of the 95\% confidence intervals (CIs) for the survival parameters, along with the corresponding CI lengths. Across all parameters, our proposed approach achieves the highest coverage probabilities, demonstrating improved reliability in interval estimation compared to unadjudicated analysis. Importantly, the improved coverage achieved by our method does not come at the cost of substantially wider intervals. The lengths of the 95\% confidence intervals under our approach are generally comparable to those obtained from the adjudicated data analysis, suggesting that the gain in coverage is not driven by overly conservative (i.e., wider) intervals.
\begin{table}[H]
	\centering
	\caption{Coverage probability with CI length comparison for the survival parameters}
	\begin{tabular}{c|c|c|c|c|c|c}
    \hline
   &\multicolumn{3}{c|}{Probability}&\multicolumn{3}{c}{95\% CI length}\\
    \hline
     Variable&Adjudication& Unadjudication& Our approach&  Adjudication& Unadjudication& Our approach\\
    \hline
     Value-BP&0.15&0.03&0.05&0.02&0.02&0.03\\
    Value-GLUC&0.93&0.95&0.99&1.81&1.75&7.99\\
     Value-CHL&0.09&0.02&0.03&0.02&0.02&0.02\\
      Base-age&0.78&0.59&0.79&0.06&0.06&0.07\\
       BMI&0.91&0.92&0.95&0.05&0.05&0.06\\
        LH&0.94&0.90&0.99&0.81&0.83&0.94\\
         AH&0.86&0.72&0.90&0.77&0.79&0.97\\
          Sex&0.86&0.62&0.86&0.65&0.66&0.78\\
          Race&0.86&0.72&0.91&0.71&0.75&0.89\\
          \hline
	\end{tabular}
	\label{tab:prob}
\end{table}
\section{Application on ARIC }
We evaluate the performance of our proposed method using data sets A and B. For both datasets, we know the true CVD death status. The results are shown in Table \ref{tab:pred}.  In the test data set (B), the percentage of true CVD death was $57\%$ among CVD death according to ICD9. This percentage is found $63\%$, which is the average probability in that group. Similarly, the predicted and the true proportion of non-CVD death are approximately 93\% and 85\%, respectively, among non-CVD death according to ICD9.  
\begin{table}[H]
   \centering
		\caption{Out-sample predictive performance on ARIC data. The posterior predictive weights are given in parenthesis.}
		\begin{tabular}{c|ccc}
			\hline
&&\multicolumn{2}{c}{ICD-9}\\
\hline
   &&Death from CVD&Death from NOT CVD\\
     \multirow{2}{*}{Adjudication}&Death from CVD&0.57 (0.63)&0.15 (0.07)\\
&Death from NOT CVD&0.43 (0.37)&0.85 (0.93)\\
   			\hline
		\end{tabular}
  \label{tab:pred}
	\end{table}
 We need to fit $KLM$ number of joint models for our approach. Based on our suggested method to find the values of K,L, and M, we choose $M=3$, $K=5$ and $L=14$ (see Appendix). For the joint model on dataset B, we assumed a proportional hazard model with baseline covariates and three features for each of the four longitudinal risk factors (SBP, DBP, GLUCOSE, and TOTCHL). The survival sub-model takes the form
\begin{eqnarray*}
				\lambda_i(t)& =& \lambda_0(t) \exp\Big[\gamma_1 \text{Baseline BMI}_i+\gamma_2\text{LH}_i+\gamma_3\text{AH}_i+\gamma_4\text{Sex}_i+\gamma_5{\text{Race}}_i+\\
                &&\hspace{3cm}\sum_{g=1}^{4}\Big(\alpha_{g1}\mu_{ig}(t)+
		\alpha_{g2}\frac{d\mu_{ig}(t)}{dt}+\alpha_{g3}\int_{t_0}^t\mu_{ig}(s)ds\Big)\Big].
		\end{eqnarray*}
A comparison of the estimates of the survival parameters using CVD death by ICD9 (unadjudicated), adjudicated CVD death, and adjusted CVD death from our approach is shown in Table (\ref{tab:comp}). Baseline BMI is an important risk factor with a positive and statistically significant effect on CVD death, as indicated by the adjudicated results. This finding matches the findings of our approach, but contrasts with the results based on CVD events from ICD-9. Additionally, individuals with education levels beyond high school (AH) experience significantly fewer CVD-related deaths compared to those with only a high school education according to both adjudicated and adjusted results by our approach. However, this trend does not hold when using ICD-9 CVD death as the true event. Furthermore, SBP demonstrates a positive and significant association with CVD deaths based on ICD-9 CVD data, while the adjudicated and adjusted CVD outcomes do not support this relationship.
\begin{table}[H]
	\centering
	\caption{Comparison between ICD-9 and adjudicated CVD for the survival parameters}
    \resizebox{\textwidth}{!}{
	\begin{tabular}{c|c|c|c|c|c|c|c|c|c}
		\hline
		&\multicolumn{3}{c|}{Unadjudicated CVD}& \multicolumn{3}{c|}{Adjudicated CVD}& \multicolumn{3}{c}{Our approach}\\
		\hline
		Variable&Estimate&2.5\% CI&97.5\% CI&Estimate&2.5\% CI&97.5\% CI&Av.Estimate&2.5\% CI&97.5\% CI\\
		\hline
		Baseline BMI&0.0139&-0.0048&0.0334&0.0260&0.0045&0.0478&0.0235&0.0220&0.0249\\
		LH&0.4293&0.2189&0.6277&0.4074&0.1701&0.6420&0.4071&0.3885&0.4257\\
		AH&-0.1609&-0.3668&0.0334&-0.3611&-0.5884&-0.1351&-0.2090&-0.2285&-0.1895\\
 Sex&0.3652&0.1611&0.5763&0.6605&0.04240&0.9053&0.5817&0.5601&0.6033\\		Race&-0.2443&-0.4469&-0.0354&-0.2512&-0.4937&-0.0138&-0.3041&-0.3268&-0.2813\\
		Value-SBP&0.0181&0.0011&0.0342&0.0120&-0.0019&0.0324&0.0202&0.0183&0.0221\\
		Slope-SBP&-0.1616&-0.6758&0.2874&-0.0116&-0.3477&0.3669&-0.0520&-0.0589&-0.0451\\
		Area-SBP&0.0000&-0.0002&0.0003&0.0001&-0.0001&0.0005&0.0000&0.0000&0.0000\\
		Value-DBP&-0.0060&-0.0332&0.0098&-0.0111&-0.0488&0.0083&-0.0258&-0.0288&-0.0228\\
		Slope-DBP&-0.3041&-1.0809&0.2494&-0.1174&-0.8201&0.2873&-0.2175&-0.2451&-0.1900\\
		Area-DBP&-0.0001&-0.0005&0.0002&0.0000&-0.0004&0.0005&0.0000&0.0000&0.0000\\
        Value-Glucose&0.0032&-0.0026&0.0105&0.0004&-0.0098&0.0090&0.0016&0.0012&0.0019\\
		Slope-Glucose&-0.4394&-1.3682&0.1490&-0.9926&-2.4080&0.0233&-0.5593&-0.6037&-0.5150\\
		Area-Glucose&0.0000&-0.0001&0.0002&0.0000&-0.0001&0.0002&0.0000&0.0000&0.0000\\
        Value-TOTCHL&-0.0010&-0.0056&0.0017&-0.0014&-0.0073&0.0013&-0.0013&-0.0016&-0.0011\\
		Slope-TOTCHL&-0.0022&-0.1282&0.1071&-0.0261&-0.2029&0.0713&-0.0206&-0.0255&-0.0156\\
		Area-TOTCHL&0.0000&-0.0002&0.0000&0.000&-0.0001&0.0001&0.0000&0.0000&0.0000\\
		\hline
	\end{tabular}}
	\label{tab:comp}
\end{table}
Our method yields the least biased estimates for 14 out of the 17 survival parameters, producing values that are closer to the adjudicated estimates than those obtained from unadjudicated data.
\section{Discussion}
In this paper, we propose a novel Bayesian approach for inference aimed at determining whether an unadjudicated event in a cohort is a true event using a cohort that contains both adjudicated and unadjudicated events. This method leverages the construction of flexible Bayesian joint models, with the addition of a Bayesian additive regression tree model to explicitly address the challenge of ICD-9 misclassification. The methodology is carefully designed to capture the complex relationship between observed data and latent true events, improving the accuracy of event identification despite misclassification inherent in the ICD-9 coding system. We provide a detailed discussion on the algorithm used for inference and the selection of appropriate priors for the model, highlighting the considerations that need to be taken into account when implementing our approach.\\
From the literature review and using data from the ARIC study, we demonstrate that the ICD-9 code suffers from significant misclassification issues, particularly in the context of identifying CVD deaths. The misclassification problem in ICD-9 codes is not an isolated issue, and for cohorts like EPESE, where adjudication is not performed, our proposed method offers a valuable tool to adjust for such misclassifications. By applying our Bayesian framework, researchers can improve the accuracy of event classification and obtain more reliable results from their data, even when definitive adjudication is unavailable.\\\\
We evaluated the performance of our approach conducting a simulation study. Our proposed method, which aims to adjust for adjudication uncertainty, yields estimates that are notably closer to those from the adjudicated data. The bias and MSE are substantially reduced compared to the unadjudicated approach, and coverage probabilities are improved, indicating more reliable inference. These findings highlight the effectiveness of our approach in mitigating the adverse effects of incomplete event adjudication.\\\\
We also applied all three approaches on ARIC data. Our analysis revealed that there are notable differences in the parameter estimation between ignoring the misclassification and correcting for the misclassification. These findings demonstrate the potential of our Bayesian approach to provide more accurate inferences, ultimately contributing to better decision-making and more reliable conclusions in epidemiological studies. The ability to adjust for ICD-9 misclassification enhances the robustness of research findings, particularly in studies where full adjudication is not feasible.\\
For cohorts like EPESE that lack event predictors and sufficient longitudinal risk factors, different imputation approaches, such as the use of multiply imputed cohort data \cite{zeki2019use,siddique2019measurement}, can improve the analyses by imputing longitudinal risk factors not directly observed within individual cohorts. The transportability assumption `automatically' holds by design in our case, as the random sample is drawn from the same study population; however, this may not hold in general settings. The validity of transported inferences relies on the transportability assumption, and violations of this assumption can lead to biased or invalid conclusions, highlighting the need for sensitivity analysis in such settings \cite{bareinboim2016causal}.\\\\
\hspace{-0.5cm}\textbf{Acknowledgments}\\
All authors were partially supported by NIH R01 HL 158963. ARIC data are available through the NHLBI BIOLINCC data repository.
\bibliography{mybib}
\end{document}